\documentclass[12pt]{article}
\usepackage{graphicx}
\usepackage[left=1in, right=1in]{geometry}

\title{Application of photon detectors in the VIP2 experiment to test the Pauli Exclusion Principle}

\author{A. Pichler$^{a}$, \vspace{-2ex} S. Bartalucci$^{b}$, \vspace{-2ex} M. Bazzi$^{b}$, \vspace{-2ex} S. Bertolucci$^{c}$, \vspace{-2ex} C. Berucci$^{a,b}$, \and M. Bragadireanu$^{b,d}$, \vspace{-2ex}  M. Cargnelli$^{a}$, \vspace{-2ex} A. Clozza$^{b}$, \vspace{-2ex} C. Curceanu$^{b,d,j}$, \vspace{-2ex} L. De Paolis$^{b}$, \and  S. Di Matteo$^{e}$, \vspace{-2ex} A. D'Uffizi$^{b}$, \vspace{-2ex} J.-P. Egger$^{f}$, \vspace{-2ex} C. Guaraldo$^{b}$, \vspace{-2ex} M. Iliescu$^{b}$, \and T. Ishiwatari$^{a}$, \vspace{-2ex} M. Laubenstein$^{g}$, \vspace{-2ex} J. Marton$^{a}$, \vspace{-2ex} E. Milotti$^{h}$, \vspace{-2ex} D. Pietreanu$^{b,d}$, \and K. Piscicchia$^{b,j}$, \vspace{-2ex} T. Ponta$^{b}$, \vspace{-2ex} E. Sbardella$^{b}$, \vspace{-2ex} A. Scordo$^{b}$, \vspace{-2ex} H. Shi$^{b}$, \vspace{-2ex} D. Sirghi$^{d}$, \and F. Sirghi$^{b,d}$, \vspace{-2ex} L. Sperandio$^{b}$, \vspace{-2ex} O. Vazquez-Doce$^{i}$, \vspace{-2ex} E. Widmann$^{a}$, \vspace{-2ex} J. Zmeskal$^{a}$}
\date{}
\begin{document}
\maketitle
\noindent \llap{$^{a}$}Stefan Meyer Institute for subatomic physics, Boltzmanngasse 3, 1090 Vienna, Austria \\
\llap{$^{b}$}INFN, Laboratori Nazionali di Frascati, C.P. 13, Via E. Fermi 40, I-00044 Frascati(Roma), Italy\\
\llap{$^{c}$}CERN, CH-1211, Geneva 23, Switzerland\\
\llap{$^{d}$}IFIN-HH, Institutul National pentru Fizica si Inginerie Nucleara Horia Hulubbei, Reactorului 30, Magurele, Romania\\
\llap{$^{e}$}Institut de Physique UMR CNRS-UR1 6251, Universit\'e de Rennes, F-35042 Rennes, France\\
\llap{$^{f}$}Institut de Physique, Universit\'{e} de Neuch\^{a}tel, 1 rue A.-L. Breguet, CH-2000 Neuch\^{a}tel, Switzerland\\
\llap{$^{g}$}INFN, Laboratori Nazionali del Gran Sasso, I-67010 Assergi (AQ), Italy\\
\llap{$^{h}$}Dipartimento di Fisica, Universit\`{a} di Trieste and INFN-Sezione di Trieste, Via Valerio, 2, I-34127 Trieste, Italy\\
\llap{$^{i}$}Excellence Cluster Universe, Technische Universit\"at M\"unchen, Boltzmannstrasse 2, D-85748 Garching, Germany\\
\llap{$^{j}$}Museo Storico della Fisica e Centro Studi e Ricerche Enrico Fermi, Piazza del Viminale 1, 00183 Roma, Italy

\begin{abstract}
The Pauli Exclusion Principle (PEP) was introduced by the austrian physicist Wolfgang Pauli in 1925. Since then, several experiments have checked its validity. From 2006 until 2010, the VIP (VIolation of the Pauli Principle) experiment took data at the LNGS underground laboratory to test the PEP. This experiment looked for electronic 2p to 1s transitions in copper, where 2 electrons are in the 1s state before the transition happens. These transitions violate the PEP. The lack of detection of X-ray photons coming from these transitions resulted in a preliminary upper limit for the violation of the PEP of $4.7 \times 10^{-29}$. Currently, the successor experiment VIP2 is under preparation. The main improvements are, on one side, the use of Silicon Drift Detectors (SDDs) as X-ray photon detectors. On the other side an active shielding is implemented, which consists of plastic scintillator bars read by Silicon Photomultipliers (SiPMs). The employment of these detectors will improve the upper limit for the violation of the PEP by around 2 orders of magnitude.
\end{abstract}

\section{Introduction}
To the best of our knowledge, there are 2 spin-separated classes of particles: fermions with half-integer spin and bosons with integer spin. The Pauli Exclusion Principle is only valid for fermions. It states that two fermions can not be in the same quantum state \cite{Pauli}. It is a fundamental principle in physics, especially were many-fermion systems are concerned. Also, an intuitive explanation is still missing \cite{Feynman}. For these reasons, a thorough test of the PEP is necessary.

In the past, a few experiments have tested the PEP. Some of them put stringent limits to the probability of the violation of the Pauli Exclusion Principle in fermionic systems, for example the experiment by the DAMA collaboration \cite{DAMA}. But some of these experiments, like the one DAMA conducted, investigated transitions of stable systems from a non-Pauli violating state to a Pauli violating state. These transitions violate the Messiah-Greenberg superselection rule \cite{MG}, which forbids the change of the symmetry state of a stable system. To bypass this rule, the VIP experiment introduces ``new'' electrons in the system by the means of an electric current. 

%As the expected number of X-rays from non-Paulian transitions is very low, it is imperative to take great care of the data taking and event selection. For this reason, two detector systems have been set up for the VIP 2 experiment, with the goal to reduce the background and select promising events.

\section{Experimental Method}
\begin{figure}[h]
 \centering
 \includegraphics[width=0.70\textwidth]{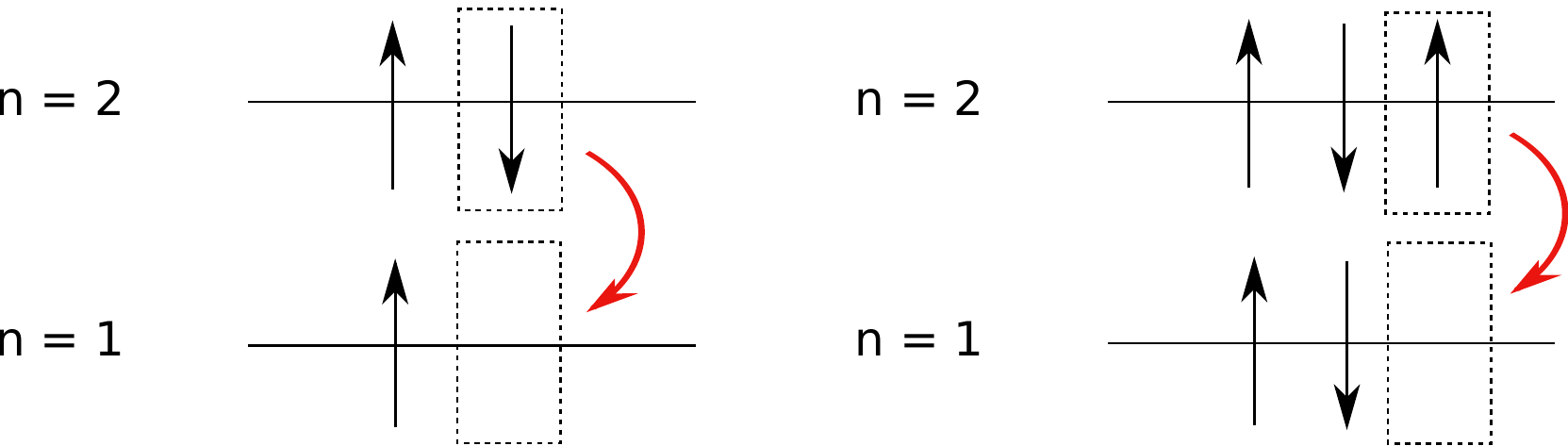}
 % energy_scheme.pdf: 479x136 pixel, 72dpi, 16.90x4.80 cm, bb=0 0 479 136
 \caption{Normal 2p to 1s transition with an energy of around 8 keV for Copper (left) and Pauli-violating 2p to 1s transition with a transition energy of around 7,7 keV in Copper (right).}
 \label{fig:energy_scheme}
\end{figure}

\begin{figure}[h]
 \centering
 \includegraphics[width=0.6\textwidth]{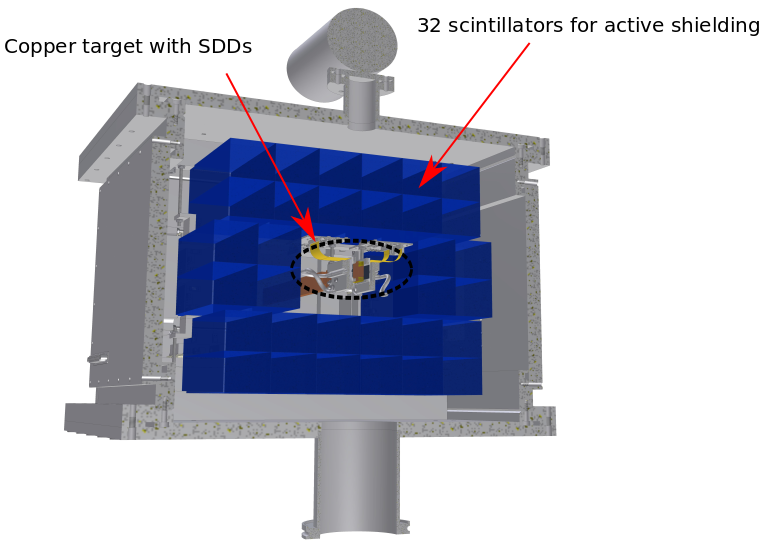}
 % active_shielding.png: 714x537 pixel, 72dpi, 25.19x18.94 cm, bb=0 0 714 537
 \caption{Schemtatic view of the VIP2 setup with the copper target and the SDDs in the center and the scintillators for active shielding around them.}
 \label{fig:exp_scheme}
\end{figure}

In the VIP2 experiment, ``new'' electrons are introduced into a copper target, which consists of 2 strips with 25 $\mu$m thickness having around 6 cm length, by an electric current of 100 A. These electrons have a certain probability to interact with and be captured by copper atoms. In the course of this interaction, the electrons might form a new symmetry state with the electrons of the atom. This process is the reason why the VIP2 experiment does not violate the Messiah-Greenberg superselection rule, because the current electrons are ``new'' to the electrons in the atom, i.e. they have no predefined symmetry state. With a certain probability which should be determined by the experiment, the newly formed symmetry state has a symmetric component and the electron can undergo transitions like that on the right side of figure \ref{fig:energy_scheme} during its cascading process.
This transition violates the PEP, as there are 2 electrons in the same state before and after it occurs. One important thing to note is that the non-Paulian transition has about 300 eV lower energy with respect to the allowed one. The energy of the X-rays which are emitted during the cascading process are recorded by Silicon Drift Detectors (SDDs), which are mounted close to the copper target. The recorded spectrum is then analyzed looking for an excess of counts over the background in the energy range of the forbidden transition, which is around 7,7 keV. The excess of counts or the lack of it will either discover a violation of the PEP or else determine a new upper bound for its violation. A schematic view of the setup is shown in figure \ref{fig:exp_scheme}.

The predecessor experiment VIP took data from 2006 to 2010 with the same measurement principle, but with Charge Coupled Devices (CCDs) as X-ray detectors. The data analysis resulted in a preliminary upper limit for the violation of the PEP of $4.7 \times 10^{-29}$ \cite{Catalina1, Catalina2}. During the currently ongoing preparation of the VIP2 experiment at the laboratory of the Stefan Meyer Institute (SMI) in Vienna, two important upgrades regarding the detector systems were conducted. On the one hand, the CCDs were replaced by Silicon Drift Detectors, which have a superior energy resolution and offer timing capability. On the other hand, an active shielding system was implemented, to reduce the background in the energy region of the forbidden transition. These systems will have an important contribution to improve the limit for the violation of the PEP by two orders of magnitude with new data which will come from the VIP2 experiment \cite{VIP_proposal}. The performance of these detectors is presented below.

\section{The Detectors of VIP2}
It is of paramount importance for an experiment with a low expected rate of interesting events such as VIP2 to carefully measure and select events. For the selection of events, an active shielding system is implemented in the VIP2 setup. For the recording of the energy of the photons, Silicon Drift Detectors are used. These two detector systems are described in what follows in more detail.

\subsection{Active Shielding System}
The active shielding system is used to select events with energy deposition in the SDDs which are not in coincidence with incoming cosmic rays or environmental background. These kinds of radiation can, on one hand, cause secondary radiation inside the setup box, through scattering effects. On the other hand, they can ionize electrons from the 1s orbital of copper atoms, which leads to an electronic X-ray transition to the K-shell. Both these effects can cause events in the SDDs, which are in the interesting energy region of around 8 keV. Events caused by these effects are not connected to non-Paulian 2p to 1s transitions, which means they are background. Consequently X-rays recorded in the SDDs which are in coincidence with events in the active shielding system are excluded from the data analysis.

\begin{figure}[h]
 \centering
 \includegraphics[width=0.90\textwidth]{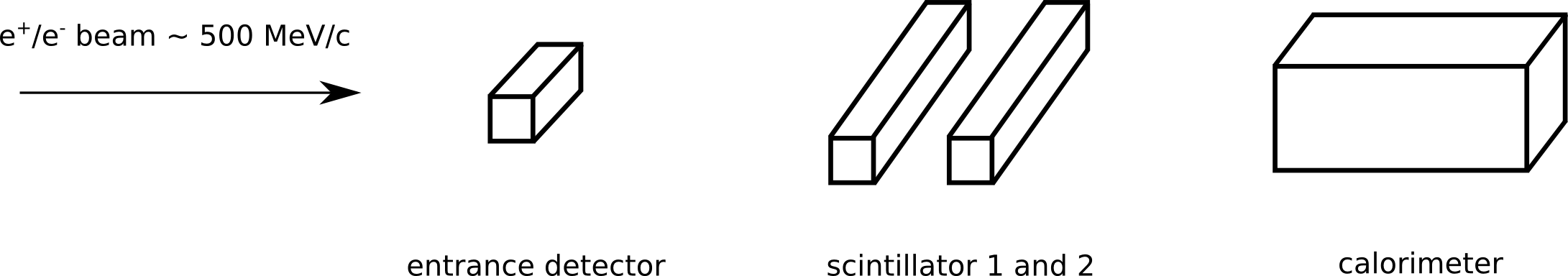}
 % BTF_scinti_test_scheme.png: 676x136 pixel, 72dpi, 23.85x4.80 cm, bb=0 0 676 136
 \caption{Scintillator and SiPM test setup scheme at the Beam Test Factory at LNF-INFN.}
 \label{fig:BTF_scheme}
\end{figure}

The veto system consists of 32 plastic scintillators of the type EJ-200 read out by pairs of Silicon Photomultipliers on one end. The SiPMs have an active area of 3x3 mm$^{2}$ each. The scintillators have the dimension of 38 mm x 40 mm x 250 mm and they cover a solid angle of over 90 \% with respect to the copper target \cite{Catalina3} (see also figure \ref{fig:exp_scheme}). To improve the light collection efficiency, the scintillators are wrapped in aluminium foil and then in black tape. This configuration of wrapped EJ-200 plastic scintillators read out on one end by two SiPMs was tested regarding the detection efficiency and the time resolution at the Beam Test Factory (BTF) at LNF-INFN. BTF-LNF delivers electrons or positrons with a momentum of 500 MeV/c. A scheme of the test setup is shown in figure \ref{fig:BTF_scheme}.
The detection efficiency for this kind of radiation was measured for two scintillators simultaneously. The impact point of the beam was varied from the scintillator side near to the SiPM readout to the far side of the SiPM readout. The trigger was made by an ``AND'' of the entrance detector and the calorimeter. To get the detection efficiency, the number of scintillator signals was compared to the number of triggers. For all impact points, the efficiency was found to be higher than 97,5 \%. With the same configuration, the scintillator time resolution was measured. It was found that the scintillator timing has a spread of 3 ns (FWHM) with respect to the calorimeter timing. The goal is to reduce the background in the energy region of interest by around one order of magnitude with the help of the active shielding.

\subsection{Silicon Drift Detectors}
Two chips of three cells of Silicon Drift Detectors (SDDs) with 100 mm$^{2}$ active area each are used in the VIP2 experiment to detect X-ray photons in the energy range from a few hundred eV to around 20 keV. The two chips are located on either side close to the copper target in order to cover as much solid angle as possible. The SDDs are cooled down to a working temperature of around 100 K by a liquid argon system.

\begin{figure}[h]
 \centering
 \includegraphics[width=0.99\textwidth]{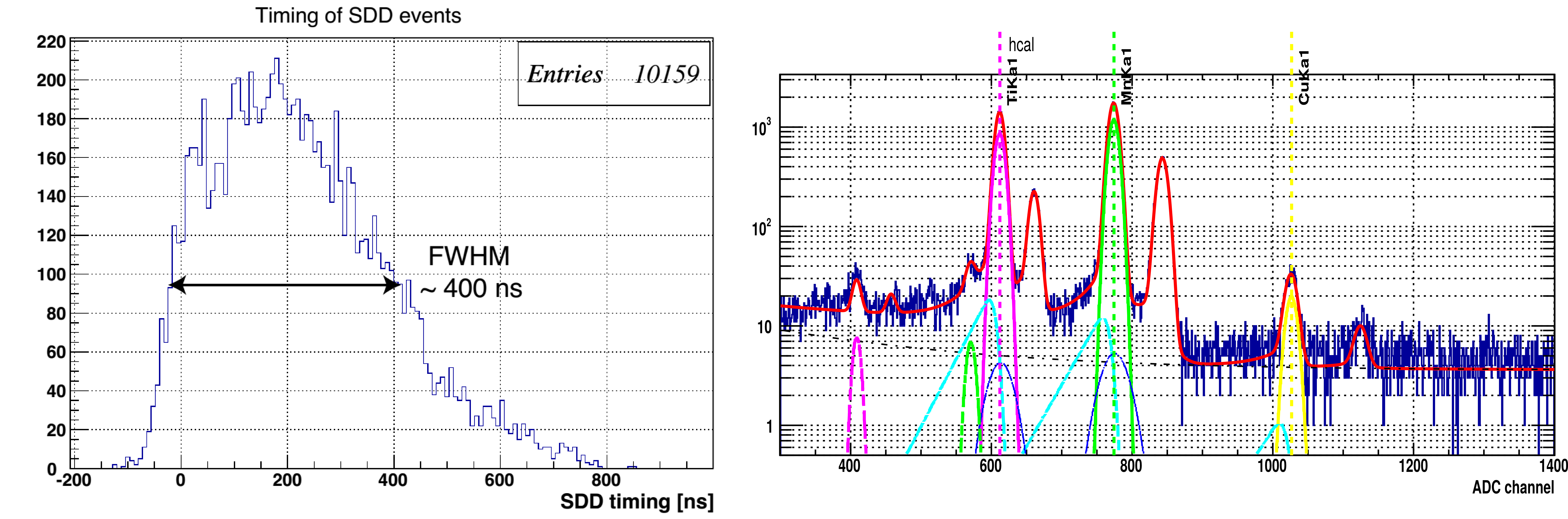}
 % timing_sdd.png: 3616x2610 pixel, 300dpi, 30.62x22.10 cm, bb=0 0 868 626
 \caption{Timing of the SDD events with respect to the scintillator timing at t = 0 ns (left). Energy spectrum of one SDD with a fit for calibration (right).}
 \label{fig:sdd_timing}
\end{figure}

When the radiation hits the SDDs, a certain number of electrons is created, which depends on the energy and the kind of the incident radiation. The electrons are guided to an anode, where they are collected and used to measure the energy of the X-ray. An advantage of using SDDs is that they provide not only energy information about the radiation, but also timing information. This enables the rejection of events which are in coincidence with scintillator events of the active shielding system. The proof that this is possible was obtained with cosmic ray triggers. The SDDs were put into the setup (see figure \ref{fig:exp_scheme}) and the SDD timing was recorded with respect to the triggers from the scintillators. The results are shown in figure \ref{fig:sdd_timing}.
The time resolution of the SDDs was found to be around 400 ns (FWHM) in this test. Taking into account the scintillator trigger rate of around 10 Hz in the laboratory at Stefan Meyer Institute, this time resolution is enough to discard SDD events caused by background events.

The energy resolution was tested in the laboratory at SMI. For this purpose, the SDDs were irradiated with a 10 $\mu$Cu Fe-55 source through a titanium calibration foil. The Ti-K$\alpha$ and the Mn-K$\alpha$ line were used to calibrate the spectrum and to calculate the energy resolution at 6 keV (Mn-K$\alpha$ line). The resolution ranged from 140 eV to 160 eV for the six tested SDDs. As an example, a spectrum which was used for the calibration of one of the SDDs is shown on the right side of figure \ref{fig:sdd_timing}. Apart from the Ti and Mn K-lines, the Cu K-lines are visible, which come from K-shell ionization through cosmic rays or radiation from the surroundings and subsequent recombination. 

VIP2 will improve the limit for the violation of the PEP by two orders of magnitude compared to the VIP result. 

\section{Outlook}
\label{sec:outlook}
In autumn 2015, the experiment was transported to the underground laboratory INFN-LNGS in Gran Sasso, where the background will be further reduced compared to the laboratory at SMI. After implementing the setup and completing stability checks in the new environment, a long term data taking of 2-3 years will start. All the mentioned successful tests make us confident to achieve an improvement of the limit of the PEP violation by two orders of magnitude, or discover a PEP violation, after this data taking period.\\ \\

{\small \textbf{Acknowledgement}
\quad We want to thank H. Schneider, L. Stohwasser and D. St\"uckler from the Stefan Meyer Institute for their important contributions for the design and the construction of the VIP2 setup and the staff of the INFN-LNGS laboratory for the support during all phases of the experiment. We acknowledge the support from the: HadronPhysics FP6 (506078), HadronPhysics2 FP7 (227431), HadronPhysics3 (283286) projects, EU COST Action 1006 (Fundamental Problems in Quantum Physics), Austrian Science Fund (FWF), which supports the VIP2 project with the grants P25529-N20 and W1252-N27 (doctoral college particles and interactions), Centro Fermi (project: Open problems in quantum mechanics). Furthermore, this paper was made possible through the support of a grant from the John Templeton Foundation (ID 581589). The opinions expressed in this publication are those of the authors and do not necessarily reflect the views of the John Templeton Foundation.}
\\ \\

\end{document}